\documentclass[12pt]{article}

\newlength{\minitwocolumn}
\usepackage{amsmath}
\usepackage{amssymb,amscd}
\usepackage{bm}
\usepackage{amsbsy}

\newcommand{\beq}{\begin{equation}}
\newcommand{\eeq}{\end{equation}}
\newcommand{\bea}{\begin{eqnarray*}}
\newcommand{\eea}{\end{eqnarray*}}
\newcommand{\beqa}{\begin{eqnarray}}
\newcommand{\eeqa}{\end{eqnarray}}

\newcommand{\bbr}{\boldsymbol{r}}
\newcommand{\hbe}{\hat{\boldsymbol{e}}}

\newcommand{\bW}{\boldsymbol{W}}
\newcommand{\bX}{\boldsymbol{X}}
\newcommand{\bY}{\boldsymbol{Y}}
\newcommand{\bZ}{\boldsymbol{Z}}

\newcommand{\vPhi}{{\boldsymbol{\Phi}}}
\newcommand{\vPsi}{{\boldsymbol{\Psi}}}
\newcommand{\del}{\partial}

\newcommand{\qed}{\nobreak \ifvmode \relax \else
      \ifdim\lastskip<1.5em \hskip-\lastskip
      \hskip1.5em plus0em minus0.5em \fi \nobreak
      \vrule height0.75em width0.5em depth0.25em\fi}

\newcommand{\be}{\boldsymbol{e}}

\begin{document}

\baselineskip 0.7cm

\begin{titlepage}
\vspace{4\baselineskip}
\begin{flushright}
\end{flushright}
\vspace{0.6cm}
\begin{center}
{\Large\bf  Gauge theory and little gauge theory
}
\end{center}
\vspace{1cm}
\begin{center}
{Kozo KOIZUMI
\footnote{E-mail: kouzou.koizumi@gmail.com}}
\end{center}
\vspace{0.2cm}
\begin{center}
{
\it Department of Physics,
Kyoto Sangyo University, \\ 
Kyoto 603-8555, Japan}
\medskip
\vskip 10mm
\end{center}
\vskip 10mm
\begin{abstract}
The gauge theory is the most important type of the field theory, in which 
the interactions of the elementary particles are described by the exchange of the gauge bosons.
In this article, the gauge theory is reexamined as geometry of the vector space, 
and a new concept of "little gauge theory" is introduced.
A key peculiarity of the little gauge theory is that the theory is able 
to give a restriction for form of the connection field. 
Based on the little gauge theory, Cartan geometry, 
a charged boson and the Dirac fermion field theory are investigated.
In particular, the Dirac fermion field theory leads to an extension 
of Sogami's covariant derivative. 
And it is interpreted that Higgs bosons are included 
in new fields introduced in this article.
\end{abstract}
\vspace{0.1cm}
\vskip 0.8cm

\end{titlepage}

\section{Introduction}

The discoveries of Higgs boson-like particle \cite{LHC:2012}\cite{CMS:2012} 
and the gravitational wave signal \cite{LIGO:2016} are very impressive 
for the elementary particle physics. And the dark matter and dark energy physics are also fascinating.
In elementary particle physics, the standard model is the most successful model to explain almost 
experimental data. In the background of the success in the standard model, 
the gauge theory plays important roles.
In the standard model the strong and electro-weak interactions of the elementary particles 
are described by the gauge symmetry SU(3)$\times$ SU${ }_{\rm L}$ (2) $\times$ U$_{\rm Y}$(1). 
However, the standard model does not involve a theory explaining the interaction of the gravity.
On the other hand, the gravity is explained by the gauge theory \cite{Utiyama:1956}.
Towards to a unified concept of their interactions and Higgs interaction, 
the gauge theory should be reexamined.
Its spirit is resembled as the idea of Sogami's generalised covariant derivative\cite{Sogami:1995}. 
At the present, the quantum theory of gravity is not established in spite of 
a lot of efforts and attempts.
It is one of the very important tasks to construct a renormalizable and unitary quantum gravity.
Also for the theoretical physicists, a unified theory of all interactions including the gravity is 
one of our dreams. 

In this article, before going to new physics beyond the standard model, 
we reinvestigate "What is the gauge theory?" in the framework of the vector analysis.
The meanings of the gauge theory are that physical reality does not depend on any artificial point
 of view. The more concrete definition of the gauge theory is given by the section 2.
In section 2, we explain the gauge theory as geometry of the vector space. 
A concept of "pull back differential" is introduced, and 
several results by the differential geometry are reproduced in the vector analysis.
In particular, the metric compatibility condition is naturally derived. 
It is not  obvious how the connection field introduced in the gauge theory are related with 
the physical quantities as Hermite fields.  
In section 3, we introduce a concept of "little gauge theory". The little gauge theory 
is mathematically the gauge theory for the constant metric frame bundle. 
The little gauge theory restricts the form of the connection field.
The relation between the little gauge theory and the gauge theory is discussed briefly.
As applications of the little gauge theory, Cartan geometry,  U(1) gauge theory of a charged boson, 
and the Dirac fermion field theory are presented in the remaining part of this article.
In section 4, based on the gauge theory and the little gauge theory, some known results of 
Cartan geometry are derived. The relations between both theories are explicitly obtained 
by the external vielbein.
In section 5, the gauge theory and little gauge theory of a charged boson are illustrated.
Then the usefulness of the little gauge theory will be realized. 
In section 6, the Dirac fermion field theory is presented. 
Based on the little gauge theory, gauge fields are introduced by 
the pull back differential and the metric compatibility condition.
It leads to an extension of Sogami's covariant derivative \cite{Sogami:1995}. 
The implications of their gauge fields are discussed.
In particular, a new interpretation of Higgs bosons is given.
In last section 7, other applications of the little gauge theory 
are discussed. By following the results of section 6, 
the constructing method of Lagrangian for the Dirac fermion field theory is discussed.
Based on the little gauge theory, some of new directions are proposed.
 
\section{Gauge theory of the vector space}

The gauge theory is just defined by the following sentence:
"{\it All physical realities in nature should not depend on 
any choice of the internal and external coordinate frames 
which are artificially introduced in order to represent the 
quantities of their physical realities.}" 

In this section, the gauge theory as the geometry of the vector space is explained and reexamined.

It is a well-known example that the abstract vectors do not 
depend  on the choices of the basis that are artificially introduced in 
order to represent the abstract vectors.
Suppose a physical reality $\bZ$ is a quantity in the $N$ 
dimensional vector space over the K field. 
Then one can artificially define the basis vectors $\be_{I}\ (I=1,\cdots N)$ 
in the vector space. 
With the basis vectors, the physical reality $\bZ$ is represented by
\beq
    \bZ=Z^{I}\be_{I},
\eeq 
where Einstein's summation rule is adopted through this article. 
The component $Z^{I}$ is called the representation of $\bZ$ under the basis $\be_{I}$, which
representation $Z^{I}$ depends on the choice of the basis vector.

If the vector space is equipped with an inner 
product \footnote{The positivity of the inner product is not demaded in this article 
in order to include indefinite inner product.}
 $(\bX, \bY)$ over the K field, we have the following relation:
\beq
    (\be_{I},\bZ)=Z^{J} E_{{I}{J}},
\eeq
where $E_{IJ}= (\be_{I},\be_{J})$. This matrix composed of $E_{{I}{J}}$ is ordinarily called the metric.
If the matrix composed of $E_{IJ}$ have the inverse matrix composed of $E^{JI}$,
$Z^{J}$ is represented by
\beq
    Z^{J}=E^{JI}(\be_{I},\bZ).
\eeq
 
For two physical realities $\bX$ and $\bY$ in the same inner product space, 
the inner product is represented by
\beq
    (\bX,\bY)=(X^{I}\be_I,Y^{J}\be_J)={X}^{\dag I}Y^{J} E_{{I}{J}},    
\eeq
where the symbol $\dag$ stands for Hermite conjugate. 

For the choice of another basis vector $\be'_{I}$, 
the physical reality $\bZ$ is reexpressed as $\bZ={Z'}^{I}\be'_{I}$. 
Any physical reality $\bZ$ should not depend on the choice of the basis.
Therefore if $\be'_{I}={\Omega_{I}}^{J}\be_{J}$, then the two components ${Z'}^{I}$ and $Z^{I}$ are related by
\beq
    Z'^{I}={\Omega^{-1\ I}}_{J}Z^{J}.
\eeq
Here the matrix composed of ${\Omega^{-1\ I}}_{J}$ is the inverse matrix of 
the matrix composed of ${\Omega_{I}}^{J}$. 
The  two metrics of $E_{IJ}=(\be_{I},\be_{J})$ and $E'_{IJ}=(\be'_{I},\be'_{J})$ have the relation:
\beq
    E'_{IJ}=E_{KL}{{{\Omega^\dag}_{I}}}^{K} {\Omega_{J}}^{L}.
\eeq

So far we did not discuss about the external coordinate, which represents the space-time points in our world.
Any physical reality should not depend on the choices of the external coordinate frame. 
On our recognition of the space-time, the space-time is regarded as the 4-dimensional manifold. 
Therefore each point of the manifold is parametrised by the general external coordinate 
$q^\mu\ (\mu=0,1,2,3)$
\footnote{It does not matter the reason why the space-time 
is described by a 4 macroscopic variable.}. 
It is assumed that the physical reality exists on 
each point of the space-time. If the physical reality is 
the abstract vector, the vector space exists on each point of the space-time manifold. 
Therefore it makes the vector bundle on the manifold. 
Supposed that $\bZ$ is a physical reality in the vector bundle. 
Then $\bZ$ is represented by the vector function of the general coordinate $q^\mu$, 
{\it i.e.} $\bZ=\bZ(q)$. 

Let us take a point P close to a point Q and let their external 
coordinates be $q^{\mu}$ and $q^{\mu}+dq^{\mu}$, respectively.
The difference of $*\bZ(q+dq)$ and $\bZ(q)$ is
\beq
      d\bZ(q)=*\bZ(q+dq)-\bZ(q),
\eeq
where $*$ means that $\bZ(q+dq)$ is evaluated on the point P. 
Mathematically $*\bZ(q+dq)$ is the pull back for the 
isometry map $f: q^\mu\to q^\mu+dq^\mu$. 

The basis vector $\be_{I}(q)$ can be defined on each point of the manifold.
The choice of the basis and the external coordinate is 
mathematically equivalent to taking the section of the vector bundle. 
Let it be written by $\bZ(q)={Z}^{I}(q)\be_{I}(q)$ on the point P. Then we obtain
\beqa
    d\bZ(q)
	&=&*({Z}^{I}(q+dq)\be_{I}(q+dq))-{Z}^{I}(q)\be_{I}(q)\nonumber\\
	&=&(*{Z}^{I}(q+dq)-Z^{I}(q))*\be_{I}(q+dq)+Z^{I}(q)(*\be_{I}(q+dq)-\be_{I}(q)).
\eeqa
Here we can evaluate the two quantities $*Z^{I}(q+dq)$ and $*\be_{I}(q+dq)$ as the followings:
\beqa
    *{Z}^{I}(q+dq)&=& Z^{I}(q)+\displaystyle\frac{\partial Z^{I}(q)}{\partial q^\mu}dq^\mu,\label{comdiff}\\
    *\be_{I}(q+dq)&=&\be_{I}(q)+{{{\cal A}_{\mu}}^{J}}_{I}(q)\be_{J}(q) dq^\mu,\label{basisdiff}
\eeqa 
where we neglected the higher order terms more than the second order terms for $dq$.
Eq.(\ref{comdiff}) means the ordinary partial differential for the component fields $Z^I(q)$.
The term ${{{\cal A}_{\mu}}^{J}}_{I}(q)$ of Eq.(\ref{basisdiff}) is the connection field \footnote{This connection field depends on the choice of the external coordinate and the basis. In our recognition, since the external coordinates is four dimensions, 
the connection field is defined only in the four dimensions.}  describing
the difference between the basis vectors $\be_{I}(q)$ and $*\be_{I}(q+dq)$.
For the simplicity, we introduce the symbol $\del_\mu$ as the followings:
\beqa
    \partial_\mu Z^{I}(q)&=&\displaystyle \lim_{dq^\mu\to 0}\frac{ *{Z}^{I}(q+dq)-Z^{{I}}(q)}{dq^\mu},\quad\\
     \partial_\mu \be_{I}(q)&=&\displaystyle \lim_{dq^\mu\to 0}\frac{ *{\be_{I}}(q+dq)-\be_{I}(q)}{dq^\mu}={{{\cal A}_{\mu}}^{J}}_{I}(q)\be_{J}(q).
\eeqa
We call the symbol $\partial_\mu$ the pull back differential, 
which is related to the parallel transport.
The Leibniz rule holds for the pull back differential.
Thus $d\bZ(q)$ is obtained as 
\beqa
    d\bZ(q)&=&(\partial_\mu Z^{I}(q)) \be_{I}(q)dq^\mu+Z^{I}(q)(\partial_\mu \be_{I}(q)) dq^\mu\nonumber\\
    &=&(\partial_\mu Z^{I}(q)) \be_{I}(q)dq^\mu+Z^{I}(q)({{{\cal A}_\mu}^{J}}_{I} \be_{J}(q)) dq^\mu.
\eeqa
Then we have the convenient notation called as the covariant derivative ${\cal D}_{\mu}$:
\beq
    d\bZ(q)
	={\cal D}_{\mu} Z^{I}(q) \be_{I}(q) dq^\mu,
\eeq
where
\beq
	{\cal D}_{\mu} Z^{I}(q)=\partial_\mu Z^{I}(q)+{{{\cal A}_{\mu}}^{I}}_{J}(q)Z^{J}(q).
\eeq

Supposed that the inner product of the vector bundle space is defined on each point of the manifold,
we can define the metric of the vector space as
\beq
    E_{IJ}(q)=(\be_{I}(q),\be_{J}(q)). \label{lmetricdef}
\eeq 
By the pull back differential of Eq.(\ref{lmetricdef}) 
\beqa
 \partial_{\mu} E_{IJ}(q)
 &=&\left(\partial_\mu \be_{I}(q),\be_{J}(q)\right)+\left(\be_{I}(q),\partial_\mu\be_{J}(q)\right)\nonumber\\
 &=&{{{{\cal A}^\dag}_{\mu}}^{K}}_{I}(q) E_{KJ}(q) +{{{\cal A}_{\mu}}^{K}}_{J}(q) E_{IK}(q)
 \label{diffmetric2},
\eeqa
we obtain the metric compatibility condition
\beq
    \partial_{\mu} E_{IJ}(q)- \left({{{{\cal A}^\dag}_{\mu}}^{K}}_{I}(q) E_{KJ}(q) 
    +{{{\cal A}_{\mu}}^{K}}_{J}(q) E_{IK}(q)\right)=0.\label{metcomcon}
\eeq
Once the generic metric $E_{IJ}(q)$ is chosen, 
the metric compatibility condition gives restriction between
the connection field and its Hermite conjugate field. 
However, it is not clear 
how the connection field can be described in terms of Hermite fields 
in the generic case. Therefore little gauge theory, for the sake of linking 
the connection field to Hermite field, 
is proposed in the next section.
 
For the different choice of the basis vector $\be'_{I}(q)$, the physical reality $\bZ(q)$ 
is represented by $\bZ(q)={Z'}^{I}(q)\be'_{I}(q)$. 
For the pull back differential of $\be'_{I}(q)$, one can define 
the new connection field ${{{{\cal A}'}_{\mu}}^{J}}_{I}(q)$ as 
\beq
    \partial_\mu \be'_{I}(q)={{{{\cal A}'}_{\mu}}^{J}}_{I}(q)\be'_J(q).\label{nconn}
\eeq
Supposed that the basis vector $ \be'_{I}(q)$ has the relation 
\beq
 \be'_{I}(q)={\Omega_{I}}^{J}(q)\be_{J}(q),\label{gtra}
\eeq 
where ${\Omega_{I}}^{J}(q)$ is assumed as the $C^{(2)}$ differentiable function. The ${\Omega_{I}}^{J}(q)$ transformation is called 
the gauge symmetric transformation of the vector space basis. The new metric $E'_{IJ}(q)$ for the basis $\be'_{I}(q)$ is related 
to the old metric for the $\be_{I}(q)$ as 
\beq
    E'_{IJ}(q)={{({\Omega^\dag}})^{K}}_{I}(q){{({\Omega}})^{L}}_{J}(q)E_{KL}(q).
\eeq
The pull back differential of Eq. (\ref{gtra}) is obtained as
\beqa
    \partial_{\mu}\be'_{I}(q)=(\partial_\mu{\Omega_{I}}^{J}(q))\be_{J}(q)
                                +{\Omega_{I}}^{J}(q)(\partial_\mu\be_{J}(q))\label{pbdep}
\eeqa
From Eqs. (\ref{nconn}) and (\ref{pbdep}), we have the following relation between ${{{{\cal A}'}_{\mu}}^{J}}_{I}(q)$ and ${{{{\cal A}}_{\mu}}^{K}}_{J}(q)$:
\beqa
  {\Omega_{J}}^{K}(q){{{{\cal A}'}_{\mu}}^{J}}_{I}(q)
    ={\Omega_{I}}^{J}(q){{{{\cal A}}_{\mu}}^{K}}_{J}(q)
       +\partial_\mu{\Omega_{I}}^{K}(q).
\eeqa
By using the inverse matrix ${\Omega^{-1I}}_{J}(q)$ of ${\Omega_{I}}^{J}(q)$, we obtain the so-called gauge transformation of the
connection field in the vector space
\beq
    {{{{\cal A}'}_{\mu}}^{J}}_{I}(q)={\Omega_{I}}^{L}(q){\Omega^{-1J}}_{K}(q) {{{{\cal A}}_{\mu}}^{K}}_{L}(q)
    + {\Omega^{-1J}}_{K}(q)\partial_{\mu}{\Omega_{I}}^{K}(q).\label{gauge1}
\eeq

Now we can define the field strength ${{{\cal F}^{I}}_{J\mu\nu}}$ of the connection field as the followings:
\beq
    {{{\cal F}^{I}}_{J\mu\nu}}\be_{I}(q)=(\del_\mu\del_\nu-\del_\nu\del_\mu)\be_{J}(q)\label{deffieldst}.
\eeq
The straightforward calculation for the right hand side of Eq. (\ref{deffieldst}) shows 
\beqa
    &&\hspace{-1cm}(\del_\mu\del_\nu-\del_\nu\del_\mu)\be_{J}(q)\nonumber\\
    &=&\del_\mu\left(\del_\nu \be_{J}(q)\right)-\del_\nu\left(\del_\mu\be_{J}(q)\right)\nonumber\\
    &=&\Big(\del_\mu{{{\cal A}_{\nu}}^{I}}_{J}(q)-\del_\nu{{{\cal A}_{\mu}}^{I}}_{J}(q)
    +{{{\cal A}_{\mu}}^{I}}_{K}(q){{{\cal A}_{\nu}}^{K}}_{J}(q)
    -{{{\cal A}_{\nu}}^{I}}_{K}(q){{{\cal A}_{\mu}}^{K}}_{J}(q)
    \Big) \be_{I}(q).
\eeqa
Therefore the field strength for the basis $\be_{I}(q)$ takes the following form
\beq
     {{{\cal F}^{I}}_{J\mu\nu}}(q)=\del_\mu{{{\cal A}_{\nu}}^{I}}_{J}(q)-\del_\nu{{{\cal A}_{\mu}}^{I}}_{J}(q)
    +{{{\cal A}_{\mu}}^{I}}_{K}(q){{{\cal A}_{\nu}}^{K}}_{J}(q)-{{{\cal A}_{\nu}}^{I}}_{K}(q){{{\cal A}_{\mu}}^{K}}_{J}(q).
\eeq
One can obtain the following relation between the field strengths $ {{{\cal F}^{I}}_{J\mu\nu}}(q)$ for the basis $\be_{I}(q)$  and $ {{{\cal F}'^{I}}_{J\mu\nu}}(q)$ for the basis $\be'_{I}(q)$:  
\beqa
    {{{\cal F}'^{I}}_{J\mu\nu}}(q)\be'_{I}(q)&=&(\del_\mu\del_\nu-\del_\nu\del_\mu)\be'_{J}(q)\nonumber\\
     &=&(\del_\mu\del_\nu-\del_\nu\del_\mu){\Omega_{J}}^{K}(q)\be_{K}(q)\nonumber\\
     &=&{\Omega_{J}}^{K}(q) {{{\cal F}^{L}}_{K\mu\nu}}\be_{L}(q).
\eeqa
Therefore under the gauge symmetric transformation (\ref{gtra}) of the vector space basis, the field strength transforms as the followings:
\beq
    {{\cal F}'^{I}}_{J\mu\nu}(q)
    ={\Omega^{-1I}}_{L}(q){\Omega_{J}}^{K}(q){{\cal F}^{L}}_{K\mu\nu}(q).
\eeq  

For another choice $\breve{q}$ of the space-time parametrisation, the physical reality $\bZ$ is represented by $\bZ(\breve{q})$. The physical reality $\bZ$ should not depend on 
the choice of the space-time parametrisation, {\it i.e.} $\bZ(q)=\bZ(\breve{q})$.
If $\breve{q}$ is the smooth function of the original space parametrisation $q$, 
$\breve{q}$ is written by $\breve{q}_\mu=\breve{q}_\mu(q)$. 
Then it is called the gauge transformation for the space-time parametrisation.
Supposed that the basis vector $\breve{\be}_I(\breve{q})$ is transformed
as $\breve{\be}_I(\breve{q})={{\breve{\Omega}}_I}{ }^J(q)\be_{I}(q)$. 
From the chain rule for the derivative, the relation between 
the pull back differential $\del_\mu$ for $q^\mu$ and  $\breve{\del}_\mu$ for $\breve{q}^\mu$ 
is given by
\beq
    {\breve{\del}_\mu}={{\Lambda_\mu}^\nu}(q){\del}_\nu,
\eeq 
where $\displaystyle{\Lambda_\mu}^\nu(q)=\frac{\del {q}^\nu}{\del \breve{q}^\mu}$. 
The connection field ${{\breve{\cal A}}_{\mu}}{{ }^{I}}_{J}(\breve{q})$ 
and the field strength ${{\breve{\cal F}}^{I}} {{ }_{J\mu\nu}}(\breve{q})$ 
for $\breve{q}$ can be defined by the pull back differential for $\breve{\be}_I(\breve{q})$, 
respectively, as
\beqa
{{\breve{\cal A}}_{\mu}}{{ }^{I}}_{J}(\breve{q})\breve{\be}_{I}(\breve{q})
&=&\breve\del_\mu\breve{\be}_{J}(\breve{q}),\nonumber\\
{{\breve{\cal F}}^{I}} {{ }_{J\mu\nu}}(\breve{q})\breve{\be}_{I}(\breve{q})
&=&(\breve\del_\mu\breve\del_\nu-\breve\del_\nu\breve\del_\mu)\breve{\be}_{J}(\breve{q}).\nonumber
\eeqa
For the gauge transformation for the space-time parametrisation, the connection field and the field strength
are transformed as
\beqa
{{\breve{\cal A}}_{\mu}}{{ }^{J}}_{I}(\breve{q})
&=&{\Lambda_{\mu}}^{\nu}(q){\breve\Omega_{I}}{ }^{L}(q)
      {\breve\Omega^{-1}}{}^J{ }_{K}(q){{{\cal A}}_{\nu}}{{ }^{K}}_{L}({q})
      +{\Lambda_{\mu}}^{\nu}(q){\breve\Omega^{-1}}{}^J{ }_{K}(q)\del_{\nu}{\breve\Omega_{I}}{ }^{K}(q),\\
{{\breve{\cal F}}^{I}} {{ }_{J\mu\nu}}(\breve{q})&=&{\breve\Omega^{-1}}{}^I{ }_{L}(q)
{\breve\Omega}_{J}{ }^K(q)
{\Lambda_{\mu}}^{\rho}(q){\Lambda_{\nu}}^{\sigma}(q){{{\cal F}}^{I}} {{ }_{J\rho\sigma}}({q}).
\eeqa
Then it turns out that the connection field does not transform as tensor in generic case.  \\

\section{Little Gauge Theory}

In the gauge theory, it is unclear how  the connection field can be described 
in terms of Hermite fields. 
Here we introduce the little gauge theory as:\\

\noindent
{\underline{\it Definition of the little gauge theory}}:\\

{\it 
Utilizing the freedom of how to choose the vector space basis, 
we can take the special basis to be the constant metric $E^f_{IJ}$ 
that is defined by $\del_{\mu}E^f_{IJ}=0$. 
The little gauge theory is the theory fixed to 
 the constant metric $E^f_{IJ}$. And the little gauge theory is 
the little subset of the gauge theory.}\\

Once the little gauge theory is constructed, the gauge theory can be brought back
by utilizing the degree of freedom of how to choose the vector space basis for physical reality.

One of the characteristics in the little gauge theory 
is to be able to restrict the form of the connection field 
by the metric compatibility condition (\ref{metcomcon}).
In fact, the metric compatibility condition for the constant metric $E^f_{IJ}$ become
\beq
    {{{{\cal A}^\dag}_{\mu}}^{K}}_{I}(q) E^f_{KJ}
    +{{{\cal A}_{\mu}}^{K}}_{J}(q) E^f_{IK}=0.
\eeq

In the little gauge theory, there still exists 
the symmetric continuous transformation ${{\Omega^f}_{I}}^{J}(q)$ 
preserving the constant metric $E^f_{IJ}$, {\it i.e.}
\beq
    E'^f_{IJ}={{\Omega^{f\dag}}_{I}}^{L}(q){{\Omega^f}_{J}}^{K}(q)E^f_{LK}=E^f_{IJ}.
\eeq 
The transformation ${{\Omega^f}_{I}}^{J}(q)$ is called the little gauge transformation.
Similar to the gauge theory, the little gauge transformation for the basis 
$\be'_{I}(q)={{\Omega^f}_{I}}^{J}(q)\be_{J}(q)$ leads 
to the transformation of the connection field and the field strength, respectively, as
\beq
    {{{{\cal A}'}_{\mu}}^{J}}_{I}(q)
    ={\Omega^f_{I}}^{L}(q){\Omega^{f-1J}}_{K}(q) {{{{\cal A}}_{\mu}}^{K}}_{L}(q)
    + {\Omega^{f-1J}}_{K}(q)\partial_{\mu}{\Omega^f_{I}}^{K}(q)\label{lgauge1}
\eeq 
and
\beq
    {{\cal F}'^{I}}_{J\mu\nu}(q)
    ={\Omega^{f-1I}}_{L}(q){\Omega^f_{J}}^{K}(q){{\cal F}^{L}}_{K\mu\nu}(q).
\eeq

Note that the little gauge theory is very similar to 
the theory referred as "the gauge theory" in the standard model and so on. 
In the remaining part of this article, the gauge theory and the little gauge theory are illustrated.
And implications of the little gauge theory will be clarified.

\section{Cartan geometry as the gauge theory and the little gauge theory}
In this section, Cartan geometry, which is an extension of Riemann geometry, is reexamined at the gauge theoretical point and the little gauge theoretical point of views. And it is shown that the external vielbein represents the map between the little gauge theory and the gauge theory.\\

\noindent
\subsection{Cartan geometry as the gauge theory} 

Let the physical reality ${\bbr}$\footnote{Note that the space-time reality {$\bbr$} is not the vector.} be the space-time reality described by the 4 dimensional manifold.
Any point P on the space-time reality $\bbr$ can be represented by the general coordinate $q^\mu$, 
{\it i.e.} $\bbr=\bbr(q)$, which is assumed as the smooth and differentiable function of $q$.
The basis vector $\hbe_\mu(q)$ of the tangent vector space T$_{\rm P}$M on P is naturally introduced by the pull back differential of $\bbr(q)$ on the point P, {\it i.e.}
\beq
    \hbe_\mu(q)=\del_{\mu}{\bbr}(q). 
\eeq
Then the tangent vector bundle TM($\bbr$) is defined over the space-time. The choice of the general coordinate $q^\mu$ and $\hbe_{\mu}(q)$ corresponds to taking the section of the tangent vector bundle TM($\bbr$). Supposed that the real inner product of the tangent vector space is defined on each space-time point. Then the metric $g_{\mu\nu}(q)$ of the basis vector is obtained as
\beq
    g_{\mu\nu}(q)=(\hbe_{\mu}(q),\hbe_{\nu}(q))\label{defgmet}
\eeq
Because of the symmetry for the real inner product, $g_{\mu\nu}(q)=g_{\nu\mu}(q)$. 
Then the inner product for the vectors $\bX(q)=X^\mu(q)\hbe_{\mu}(q)$ and $\bY(q)=Y^\mu(q)\hbe_{\mu}(q)$ in the tangent vector space is
\beq
    (\bX(q),\bY(q))=X^\mu(q)Y^\nu(q) g_{\mu\nu}(q).
\eeq

The affine connection field ${\Gamma^\lambda}_{\nu\mu}(q)$ is defined by the pull back differential of $\hbe_{\mu}(q)$ as
\beq
    \del_{\mu}\hbe_\nu(q)={\Gamma^\lambda}_{\nu\mu}(q)\hbe_\lambda(q).\label{defconne}
\eeq
For the vector $\bZ(q)=Z^\mu(q)\hbe_{\mu}(q)$ in the tangent vector, the pull back differential leads to
\beqa
    \del_{\mu}\bZ(q)&=&\Big(\del_\mu Z^\nu(q)\Big)\hbe_{\nu}(q)+Z^\nu(q)\Big(\del_\mu\hbe_{\nu}(q)\Big)\nonumber\\
    &=&\Big(\del_\mu Z^\nu(q)+{\Gamma^{\nu}}_{\lambda\mu}(q)Z^\lambda(q)\Big)\hbe_{\nu}(q).\label{pbdbZ}
\eeqa
Therefore the covariant derivative $\nabla_\mu$ of the contravariant component $Z^{\nu}(q)$ is obtained   as
\beq
    \nabla_\mu Z^\nu(q)=\del_\mu Z^\nu(q)+{\Gamma^{\nu}}_{\lambda\mu}(q)Z^\lambda(q),
\eeq
which is just interpreted as the representation for the pull back differential of $\del_\mu\bZ(q)$.
The metric compatibility condition is derived by the pull back differential of Eq.(\ref{defgmet}):
\beqa
    \del_{\mu}g_{\alpha\beta}(q)&=&(\del_{\mu}\hbe_{\alpha}(q),\hbe_{\beta}(q))
    +(\hbe_{\alpha}(q),\del_{\mu}\hbe_{\beta}(q))\nonumber\\
    &=&{\Gamma^\rho}_{\alpha\mu}(q)g_{\rho\beta}(q)+{\Gamma^\rho}_{\beta\mu}(q)g_{\alpha\rho}(q).
    \label{gmetcom}
\eeqa
Thus in this formulation this condition is not the postulation.
 
For the inverse metric  $g^{\mu\nu}(q)$ of the metric $g_{\mu\nu}(q)$, 
the pull back differential of $g^{\mu\nu}(q)$ is obtained 
from $g^{\mu\nu}(q)g_{\nu\rho}(q)=\delta^{\mu}_{\rho}$, where $\delta^\mu_{\nu}$ is the unit matrix element.
The direct calculation shows 
\beq
    \partial_{\mu}g^{\alpha\beta}(q)=-{\Gamma^{\beta}}_{\rho\mu}(q)g^{\alpha\rho}(q)
    -{\Gamma^{\alpha}}_{\rho\mu}(q)g^{\rho\beta}(q).\label{invmetcon}
\eeq
Here let introduce the covariant component $Z_{\mu}(q)$ and the dual basis $\hbe^\mu(q)$ by using $g_{\mu\nu}(q)$ and $g^{\mu\nu}(q)$, respectively, as
\beq
    Z_{\mu}(q)=g_{\mu\nu}(q)Z^{\nu}(q),\quad \hbe^{\mu}(q)=g^{\mu\nu}(q)\hbe_{\nu}(q).
\eeq
By using Eqs. (\ref{defconne}) and (\ref{invmetcon}), we can obtain the  pull back differential of $\hbe^\mu(q)$ as follows:
\beq
    \partial_{\mu} \hbe^\nu(q)=-{\Gamma^{\nu}}_{\lambda\mu}(q)\hbe^\lambda(q).
\eeq
Because the vector $\bZ(q)$ in the tangent vector is rewritten by $\bZ(q)=Z_{\mu}(q)\hbe^\mu(q)$,
the pull back differential of $\bZ(q)$ is obtained by the similar calculation to Eq. (\ref{pbdbZ}) as
\beq
    \partial_{\mu}\bZ(q)=\Big(\del_{\mu}Z_{\nu}(q)-{\Gamma^{\lambda}}_{\nu\mu}(q)Z_{\lambda}(q)\Big)\hbe^\nu(q).
\eeq
The covariant derivative $\nabla_\mu$ of the covariant component $Z_{\nu}(q)$ is defined by
\beq
    \nabla_\mu Z_\nu(q)=\del_\mu Z_\nu(q)-{\Gamma^{\lambda}}_{\nu\mu}(q)Z^\lambda(q).
\eeq
Let us consider a physical reality $\bW(q)$ in the direct product space of 
the basis $\hbe_{\mu}(q)$ and the dual basis $\hbe^{\mu}(q)$, {\it i.e.}
\beq
 \bW(q)={W^{\mu_1\cdots\mu_N}}_{\nu_1\cdots\nu_M}(q) 
    \hbe_{\mu_1}(q)\otimes\cdots\otimes\hbe_{\mu_N}(q)\otimes
    \hbe^{\nu_1}(q)\otimes\cdots\otimes\hbe^{\nu_M}(q).
\eeq
Then we define the covariant derivative in terms of the pull back differential as follows:
\beq
    \Big(\nabla_{\mu} {W^{\mu_1\cdots\mu_N}}_{\nu_1\cdots\nu_M}(q)\Big) 
    \hbe_{\mu_1}(q)\otimes\cdots\otimes\hbe_{\mu_N}(q)\otimes
    \hbe^{\nu_1}(q)\otimes\cdots\otimes\hbe^{\nu_M}(q)=\del_{\mu} \bW(q),
\eeq
where by the direct calculation it is shown that
\beqa
    \hspace{-1cm}\nabla_{\mu} {W^{\mu_1\cdots\mu_N}}_{\nu_1\cdots\nu_M}(q)
    &\!\!\!=\!\!\!&\del_{\mu}{W^{\mu_1\cdots\mu_N}}_{\nu_1\cdots\nu_M}(q)\nonumber\\
    &&+\left({\Gamma^{\mu_1}}_{\rho\mu}(q){W^{\rho\cdots\mu_N}}_{\nu_1\cdots\nu_M}(q)
      +\cdots+{\Gamma^{\mu_N}}_{\rho\mu}(q){W^{\mu_1\cdots\rho}}_{\nu_1\cdots\nu_M}(q)\right)\nonumber\\
   &&-\left({\Gamma^{\rho}}_{\nu_1\mu}(q){W^{\mu_1\cdots\mu_N}}_{\rho\cdots\nu_M}(q)
      +\cdots+{\Gamma^{\rho}}_{\nu_M\mu}(q){W^{\mu_1\cdots\mu_N}}_{\nu_1\cdots\rho}(q)\right).
      \label{extcov}
\eeqa
With this covariant derivative, the metric compatibility condition (\ref{gmetcom}) is written by
\beq
    \nabla_{\mu} g_{\nu\rho}(q)=0.
\eeq

The torsion ${T_{\mu\nu}}^\lambda(q)$ is defined as
\beq
    {T_{\mu\nu}}^\lambda(q)\hbe_{\lambda}=(\del_{\mu}\del_{\nu}-\del_{\nu}\del_{\mu}) \bbr(q)    
    \label{deftorsion}.
\eeq 
Therefore we obtain   
\beqa
     {T_{\mu\nu}}^\lambda(q)\hbe_{\lambda}(q)
    &=&\del_{\mu}\hbe_{\nu}(q)-\del_{\nu}\hbe_{\mu}(q)\nonumber\\
    &=&\Big({\Gamma^\lambda}_{\nu\mu}(q)-{\Gamma^\lambda}_{\mu\nu}(q)\Big)\hbe_{\lambda}(q).
    \label{deftorsion1}
\eeqa
If the integrability condition $(\del_{\mu}\del_{\nu}-\del_{\nu}\del_{\mu}) \bbr(q)=0$ is imposed, then the affine connection ${\Gamma^\lambda}_{\nu\mu}(q)$ is symmetric for exchange of the index $\mu$ and $\nu$. Such torsionless geometry is called Riemann geometry. However there is no theoretical reason why such imposition is demanded. 

The field strength called as the curvature is defined by
\beq
    {R^\alpha}_{\beta\mu\nu}(q)\hbe_{\alpha}(q)=(\del_{\mu}\del_{\nu}-\del_{\nu}\del_{\mu})\hbe_{\beta}(q)\label{defcur}.
\eeq
By the definition, we obtain
\beqa    
    \hspace{-0.8cm}{R^\alpha}_{\beta\mu\nu}(q)\hbe_{\alpha}(q)
    &=&\del_{\mu}\Big(\del_{\nu}\hbe_{\beta}(q)\Big)-(\mu\leftrightarrow\nu)\nonumber\\
   &=&\Big(\del_\mu{\Gamma^\alpha}_{\beta\nu}(q)-\del_\nu{\Gamma^\alpha}_{\beta\mu}(q)
  + {\Gamma^\sigma}_{\beta\nu}(q){\Gamma^\alpha}_{\sigma\mu}(q)
  -{\Gamma^\sigma}_{\beta\mu}(q){\Gamma^\alpha}_{\sigma\nu}(q)\Big)\hbe_\alpha.
\eeqa
The curvature satisfy the so-called first and second Bianchi identity.
The proof is given in the appendix.

For another choice of the external coordinate $\breve{q}^\mu$, the basis vector $\breve{\hbe}_{\mu}(\breve{q})$, the affine connection field ${\breve{\Gamma}^\lambda}{ }_{\nu\mu}(\breve{q})$, the torsion ${\breve{T}_{\mu\nu}}{ }^\lambda(\breve{q})$ and the curvature ${\breve{R}^{\alpha}}{ }_{\beta\mu\nu}(\breve{q})$ are defined in the same way as the above. Supposed that $\breve{q}^\mu$ is a smooth  function of the original coordinate $q^\mu$, {\it i.e.} $\breve{q}^\mu=\breve{q}^\mu(q)$. Then the basis vector
$\breve{\hbe}_{\mu}(\breve{q})$ is transformed as
\beq
    \breve{\hbe}_{\mu}(\breve{q})={\Lambda_{\mu}}^{\alpha}(q)\hbe_{\alpha}(q),\label{spgauge}
\eeq
where ${\Lambda_{\mu}}^{\alpha}(q)=\displaystyle\frac{\del q^\alpha}{\del \breve{q}^\mu}$.
Then by some calculations, it is shown that
the affine connection field, the torsion, and the curvature are also transformed as 
\beqa
    {\breve{\Gamma}^\lambda}{}_{\nu\mu}(\breve{q})
      &=& {\Lambda^\lambda}_{\gamma}(q){\Lambda_{\nu}}^\alpha(q){\Lambda_{\mu}}^\beta(q)
      {\Gamma^\gamma}_{\alpha\beta}(q)+{\Lambda^{\lambda}}_\gamma(q){\Lambda_{\mu}}^\alpha(q)\del_\alpha{\Lambda_{\nu}}^\gamma(q),\nonumber\\
    {\breve{T}_{\mu\nu}}{}^\lambda(\breve{q})&=&{\Lambda_{\mu}}^\alpha(q){\Lambda_{\nu}}^\beta(q){\Lambda^\lambda}_\gamma(q){T_{\alpha\beta}}^\gamma(q),\nonumber\\
    {\breve{R}^{\alpha}}{ }_{\beta\mu\nu}(\breve{q})&=&{\Lambda^\alpha}_\lambda(q){\Lambda_{\beta}}^\gamma(q){\Lambda_{\mu}}^\delta(q){\Lambda_{\nu}}^\sigma(q){R^\lambda}_{\gamma\delta\sigma}(q),
\eeqa
where the matrix elment ${\Lambda^\lambda}_{\gamma}(q)$ is the inverse matrix element of the matrix composed of  ${\Lambda_\gamma}^{\lambda}(q)$.
Note that the affine connection field does not transform as tensor, 
which reason is that the basis is transformed as Eq. (\ref{spgauge}) under the gauge transformation
of the space-time parametrisation.
In the case of the little gauge theory, 
the little connection field is transformed as tensor, which will be shown below. \\

\noindent
\subsection{Cartan geometry as the little gauge theory} 

From now, Cartan geometry as the little gauge theory is examined. 
In the little gauge theory, as the basis of the tangent vector space, 
we can take the new basis $\hbe_{a}(q)\ (a=0,1,2,3)$ so that the constant metric is
\beq
    (\hbe_a(q),\hbe_b(q))=\eta_{ab},
\eeq
where the small Latin index stands for Lorentzian index and $\eta_{ab}$ is Lorentzian metric.
The old basis and the new basis are related by the external vielbein ${e_{\mu}}^a(q)$ as
\beq
    \hbe_{\mu}(q)={e_{\mu}}^a(q)\hbe_{a}(q).\label{cLGG}
\eeq

The pull back differential of the basis vector $\hbe_{a}(q)$ leads to the introduction of the new  little connection field ${{{\cal A}_\mu}^b}_a(q)$ as
\beq
    \del_{\mu}\hbe_{a}(q)={{{\cal A}_\mu}^b}_a(q)\hbe_{a}(q).
\eeq
By the metric compatibility condition for the constant metric, the restriction for the form of 
the little connection field ${{{\cal A}_\mu}^b}_a(q)$ is derived as follows:
\beq
    \del_{\mu}\eta_{ab}={{{\cal A}_\mu}^c}_a(q)\eta_{cb}+{{{\cal A}_\mu}^c}_b(q)\eta_{ac}=0.
\eeq
As usual, it is adopted that the lowering and raising of the indices are performed
by the metric $\eta_{ab}$ and the inverse metric $\eta^{ab}$. Then
the above equation is written by
\beq
    {{\cal A}_\mu}_{ba}(q)+{{\cal A}_\mu}_{ab}(q)=0.
\eeq 
It turns out that the little connection field is antisymmetric with respect to the indices $a$ and $b$.

The dual basis is defined as $\hbe^{a}(q)=\eta^{ab}\hbe_{b}(q)$.
The pullback differential of the dual basis $\hbe^a(q)$ is
\beq
    \del_{\mu}\hbe^a(q)=\eta^{ab}\del_{\mu}\hbe_{b}(q)={{{\cal A}_\mu}_c}^a(q)\hbe^c(q)
    =-{{{\cal A}_\mu}^a}_c(q)\hbe^c(q).
\eeq
For a physical reality $\bW(q)$ in the direct product space, supposed that it is written by
\beq
    \bW(q)
    ={W^{a_1\cdots a_N}}_{b_1\cdots b_M}(q)\hbe_{a_1}(q)\otimes\be_{a_N}(q)\otimes\hbe^{b_1}(q)\otimes\hbe^{b_N}(q).
\eeq 
The covariant derivative $\nabla_\mu$ for the local Lorentzian space is defined by the pull back differential of $\bW(q)$ as
\beq
    \Big(\nabla_{\mu}{W^{a_1\cdots a_N}}_{b_1\cdots b_M}(q)\Big)\hbe_{a_1}(q)\otimes\be_{a_N}(q)\otimes\hbe^{b_1}(q)\otimes\hbe^{b_N}(q)=\partial_{\mu}\bW(q),
\eeq
which calculation shows that
\beqa
    \hspace{-1cm}\nabla_{\mu}{W^{a_1\cdots a_N}}_{b_1\cdots b_M}(q)
    &\!\!\!=\!\!\!&\del_{\mu}{W^{a_1\cdots a_N}}_{b_1\cdots b_M}(q)\nonumber\\
    &&+\Big({{{\cal A}_\mu}^{a_1}}_\rho(q){W^{\rho\cdots a_N}}_{b_1\cdots b_M}(q)+\cdots
    +{{{\cal A}_\mu}^{a_N}}_\rho(q){W^{a_1\cdots \rho}}_{b_1\cdots b_M}(q)\Big)\nonumber\\
    &&-\Big({{{\cal A}_\mu}^{\rho}}_{b_1}(q){W^{a_1\cdots a_N}}_{\rho\cdots b_M}(q)+\cdots
    +{{{\cal A}_\mu}^{\rho}}_{b_M}(q){W^{a_1\cdots a_N}}_{b_1\cdots \rho}(q)\Big).\label{Lorcov}
\eeqa
  
The field strength ${{\cal F}^a}_{b\mu\nu}(q)$ is defined by
\beq
   {{\cal F}^a}_{b\mu\nu}(q)\hbe_a(q)=\Big(\del_{\mu}\del_{\nu}-\del_{\nu}\del_{\mu}\Big)\hbe_{b}(q),
\eeq 
which leads to the following equation:
\beqa
    \hspace{-1cm}{{\cal F}^a}_{b\mu\nu}(q)\hbe_a(q)&=&\del_{\mu}\Big({{{\cal A}_{\nu}}^c}_b(q)\hbe_{c}(q)\Big)-(\mu\leftrightarrow \nu)\nonumber\\
     &=&\Big(\del_\mu{{{\cal A}_{\nu}}^a}_b(q)-\del_\nu{{{\cal A}_{\mu}}^a}_b(q)
     +{{{\cal A}_{\mu}}^a}_c(q){{{\cal A}_{\nu}}^c}_b(q)
     -{{{\cal A}_{\nu}}^a}_c(q){{{\cal A}_{\mu}}^c}_b(q)
     \Big)\hbe_{a}(q).
\eeqa

As the little gauge transformation, we can take the local Lorentzian transformation $\hbe'_a(q)={\Omega_{a}}^b(q)\be_b(q)$. The definition of the local Loretzian transformation is 
the transformation preserving the Lorentzian metric, {\it i.e.}
\beq
    \eta'_{ab}=(\hbe'_{a}(q),\hbe'_{b}(q))={\Omega_{a}}^{c}(q){\Omega_{b}}^{d}(q)\eta_{cd}=\eta_{ab}.
\eeq
Under the basis $\hbe'_a(q)$, the external vielbein ${e'_{\mu}}^a(q)$, 
the little connection field ${{{\cal A}'_\mu}^d}_a(q)$ and the field strength ${{\cal F}'^a}_{b\mu\nu}(q)$ are defined in the same way as the above.
Following the the local Lorentzian transformation, they transform as
\beqa
    {e'_{\mu}}^a(q)&=&{\Omega^a}_b(q){e_{\mu}}^b(q),\nonumber\\
    {{{\cal A}'_\mu}^d}_a(q)&=&{\Omega_a}^b(q){\Omega^d}_c(q){{{\cal A}_{\mu}}^c}_b
    +{\Omega^d}_b(q)\del_\mu{\Omega_a}^b(q),\nonumber\\
    {{\cal F}'^a}_{b\mu\nu}(q)&=&{\Omega^a}_c(q){\Omega_b}^d(q) {{\cal F}^c}_{d\mu\nu}(q),
\eeqa
where ${\Omega^a}_b(q)$ is the inverse transformation ${\Omega_{a}}^b(q)$.

For changing the external coordinate $q^\mu$ to $\breve{q}^\mu$, the basis vector transforms as $\breve{\hbe}_{\mu}(\breve{q})={\Lambda_{\mu}}^\nu(q)\hbe_{\nu}(q)$. 
Here ${\Lambda_{\mu}}^\nu(q)=\frac{\del q^\mu}{\del \breve{q}^\nu}$. 
By the direct calculations, the external vielbein, the little connection field and the field strength are also transformed as 
\beqa
     {\breve{e}_{\mu}}^a(\breve{q})&=&{\Lambda_\mu}^\nu(q){e_{\nu}}^a(q),\nonumber\\
    {{\breve{\cal A}_\mu}}{}^b_a(\breve{q})&=&{\Lambda_\mu}^\nu(q){{{\cal A}_{\nu}}^b}_a(q),\nonumber\\
    {{\breve{\cal F}^b}}{}_{a\mu\nu}(\breve{q)}
    &=&{\Lambda_\mu}^\sigma(q){\Lambda_\nu}^\rho(q){{\cal F}^b}_{a\sigma\rho}(q),
\eeqa
under the general coordinate gauge transformation.
All of their fields transform as tensor under the transformation.

The little gauge theory of Cartan geometry is related to the gauge theory of Cartan geometry 
through the external vielbein ${e_{\mu}}^a(q)$ as Eq. (\ref{cLGG}). 

The mixed direct product space by the mixed basis $\hbe_{\mu}(q), \hbe^{\nu}(q), \hbe_{a}(q)$ and $\hbe^{a}(q)$ is also considered. By the pull back differential of physical reality in the mixed direct product space, the covariant derivative $\nabla_\mu$ 
is defined as the extension of Eqs. (\ref{extcov}) and (\ref{Lorcov}).

From now, the metric, the torsion, the affine connection and the curvature obtained in the gauge theory
are related to the little connection field in terms of ${e_{\mu}}^a(q)$. 
From the pull back differential of Eq. (\ref{cLGG}) and the affine connection, we obtain
\beqa
    {\Gamma^{\lambda}}_{\nu\mu}(q)\hbe_{\lambda}(q)&=&\del_\mu\hbe_{\nu}(q)\nonumber\\
    &=&\del_{\mu}\Big({e_{\nu}}^a(q)\hbe_a(q)\Big)\nonumber\\
    &=&\Big(\del_{\mu}{e_{\nu}}^a(q)+{{{\cal A}_{\mu}}^a}_b(q) {e_{\nu}}^b(q)\Big)\hbe_{a}(q).
\eeqa
Therefore the affine connection field is related to the the little connection field as
\beq
    {\Gamma^{\lambda}}_{\nu\mu}(q){e_\lambda}^a(q)
    =\del_{\mu}{e_{\nu}}^a(q)+{{{\cal A}_{\mu}}^a}_b(q) {e_{\nu}}^b(q),
\eeq
which is rewritten in terms of the covariant derivative as
\beq
    \nabla_\mu {e_{\nu}}^a(q)=0. \label{vielcon}
\eeq
Although the above equation is called "{\it the vielbein postulate}", it is naturally derived in our formalism. Thus we call Eq.(\ref{vielcon}) the external vielbein condition. 
Similar calculations of the above to the torsion and curvature show that 
\beq
     \hspace*{-0.5cm}{T_{\mu\nu}}^\lambda(q) {e_\lambda}^{a}(q)\hbe_a(q)=
    \Big(\del_\mu {e_\nu}^a(q)+{{{\cal A}_\mu}^a}_b(q){e_\nu}^b(q)-
    \del_\nu {e_\mu}^a(q)-{{{\cal A}_\nu}^a}_b(q){e_\mu}^b(q)\Big)\hbe_a(q),
\eeq
and
\beqa
    {R^\alpha}_{\beta\mu\nu}(q){e_{\alpha}}^a(q)\hbe_a(q)
       &=&{e_\beta}^b(q)\Big(\del_\mu{{{\cal A}_\nu}^a}_b(q)
       -\del_\nu{{{\cal A}_\mu}^a}_b(q)\nonumber\\
       &&\hspace{3cm}+{{{\cal A}_{\nu}}^b}_c(q){{{\cal A}_\mu}^a}_c(q)
       -{{{\cal A}_{\mu}}^b}_c(q){{{\cal A}_\nu}^a}_c(q)\Big)\hbe_a(q)\nonumber\\
       &=&{e_\beta}^b(q){{\cal F}^a}_{b\mu\nu}(q)\hbe_a(q),
\eeqa
respectively. Note that the curvature is proportional to the field strength ${{\cal F}^a}_{b\mu\nu}(q)$ that does not involve the external vielbein. Thus the curvature is made from the little connection field and the external vielbein is not so related for the existence of the curvature.  

By the definition (\ref{defgmet}) of the metric $g_{\mu\nu}(q)$, it is induced by the following equation
\beq
    g_{\mu\nu}(q)=(\hbe_{\mu}(q),\hbe_{\nu}(q))
    =({e_{\mu}}^a(q)\hbe_{a}(q),{e_{\nu}}^b(q)\hbe_{a}(q))={e_{\mu}}^a(q){e_{\nu}}^b(q)\eta_{ab}.
\eeq
The pull back differential of the above equation leads to the following equation
\beqa
     \del_\mu g_{\nu\lambda}(q)
     &=&(\del_\mu \hbe_\nu(q),\hbe_\lambda(q))
           + (\hbe_\nu(q),\del_\mu\hbe_\lambda(q))     \nonumber\\ 
     &=&\Big(\del_\mu {e_\nu}^a(q)+{{{\cal A}_\mu}^a}_c(q){e_\nu}^c(q)\Big){e_\lambda}^b(q)\eta_{ab}\nonumber\\
    &&\quad+{e_\nu}^a(q)\Big(\del_\mu {e_\lambda}^b(q)+{{{\cal A}_\mu}^b}_c(q){e_\lambda}^c(q)\Big)\eta_{ab}.
\eeqa
By using the covariant derivative, the above equation is written by
\beqa
    \nabla_\mu g_{\nu\lambda}(q)&=&\del_\mu g_{\nu\lambda}(q)
    -\Gamma_{\nu\mu}^{\sigma}(q)g_{\sigma\lambda}(q)
    -\Gamma_{\lambda\mu}^{\sigma}(q)g_{\nu\sigma}(q)\\
    &=&\Big(\nabla_\mu{e_\nu}^a(q)\Big){e_\lambda}^b(q)\eta_{ab}
    +{e_\nu}^a(q)\Big(\nabla_\mu{e_\lambda}^b(q)\Big)\eta_{ab}=0.
\eeqa
Therefore the metric compatibility condition and the external vielbein condition are compatible 
with each other.

By the above construction, the external vielbein maps from the little gauge theory to the gauge theory.  
The most important question is "What is the external vielbein?". It is not the connection field.
However, the external vielbein is very important, because it connects 
the external coordinate and local Lorentzian frame and it also gives the definition of the distance.
In section 7, a way of solutions for the question is suggested. The full answer is beyond the scope of this article.

\section{U(1) gauge theory of the charged boson}
The most well-known example of the gauge theory is the U(1) gauge theory.
In this section, the gauge theory and the little gauge theory of the charged boson are considered.

First, we consider the charged bosonic vector field $\vPhi$ in the one dimensional complex vector space.
Supposed that the space-time point is parametrised by the general coordinate $q^\mu$.
Then the vector field $\vPhi$ is assumed as the function $\vPhi(q)$ of the space-time point.
The inner product $(V,V')$ of the complex vector fields $V$ and $V'$ is defined 
over the complex number field. For the one dimensional complex vector space over the space-time, 
in the gauge theory one can generally choose the basis $\be(q)$ with the metric $\zeta(q)=(\be(q),\be(q))=|\varphi(q)|^2$, 
where $\varphi(q)$ is a complex-valued function.  The vector field $\vPhi(q)$ is represented by 
\beq
    \vPhi(q)=\Phi(q)\be(q)
\eeq  
under the basis vector $\be(q)$. By the pull back differential of the basis, the connection field is defined as
\beq
       \displaystyle{\partial_\mu}\be(q)={\cal A}_\mu(q)\be(q).
\eeq 
Then the metric compatibility condition is 
\beq
   \del_\mu|\varphi(q)|^2=\Big({\cal A}^\dag_\mu(q)+{\cal A}_\mu(q)\Big)|\varphi(q)|^2.
\eeq
The field strength ${\cal F}_{\mu\nu}(q)$ is defined by
\beq
    {\cal F}_{\mu\nu}(q)\be(q)=\Big(\del_{\mu}\del_{\nu}-\del_{\nu}\del_{\mu}\Big)\be(q),
\eeq
which leads to
\beq
    {\cal F}_{\mu\nu}(q)=\del_\mu {\cal A}_{\nu}(q)-\del_\nu {\cal A}_{\mu}(q).
\eeq

For the gauge transformation $\be'(q)=\Omega(q)\be(q)$, the metric, the the connection field and the field strength are transformed as
\beqa
    \zeta'(q)&=&\Omega^\dag(q)\Omega(q)\zeta(q),\nonumber\\
    {\cal A}'_{\mu}(q)&=&{\cal A}_{\mu}(q)+\Omega^{-1}(q)\del_{\mu}\Omega(q),\nonumber\\
    {\cal F}'_{\mu\nu}(q)&=&{\cal F}_{\mu\nu}(q). 
\eeqa
Here note that $\Omega(q)$ is complex-valued function.
The connection field cannot remain as Hermite field 
by the degree of freedom of the gauge transformation in the gauge theory.
The physical meaning of the connection field is not explicit. 
Therefore the connection field in the gauge theory is not tamed.

In the little gauge theoretical point of view, the model of the charged boson is considered below.
This is mathematically the same as the gauge theory for the unitary frame bundle on Hermitian target manifold.
One can find the basis vector ${\tilde\be}(q)$ with the constant metric $\kappa=(\tilde\be(q),\tilde\be(q))=1$.
The vector field $\vPhi(q)$ is represented by 
\beq
    \vPhi(q)={\tilde\Phi}(q){\tilde\be}(q).
\eeq  
The pull back differential of $\tilde\be(q)$ is described by the connection field
\beq    
    \displaystyle{\partial_\mu}\tilde\be(q)=\tilde{\cal A}_\mu(q)\tilde\be(q).
\eeq
Since the metric $\kappa$ is constant, the condition for the connection field is obtained as
\beq
    \tilde{\cal A}^\dag_\mu(q)+\tilde{\cal A}_\mu(q)=0
\eeq
from the metric compatibility condition. By introducing Hermite field ${\cal B}_\mu(q)$ called as the U(1) gauge boson field, 
the field $\tilde{\cal A}_\mu(q)$ can be described by 
\beq
    \tilde{\cal A}_\mu(q)=ig {\cal B}_{\mu}(q),
\eeq
where $g$ is a coupling constant.
Then the covariant derivative of the charged boson is written by
\beq
    {\cal D}_\mu \tilde\Phi(q)=(\partial_\mu +ig {\cal B}_\mu(q))\tilde\Phi(q).
\eeq
By the (little) gauge transformation $\tilde\be'(q)=e^{i\theta(q)}\tilde\be(q)$ 
preserving the constant metric $\kappa$,
the vector field $\vPhi(q)$ is also represented by 
\beq
    \vPhi(q)={\tilde\Phi'}(q){\tilde\be'}(q),
\eeq  
where $\tilde\Phi'(q)=e^{-i\theta(q)}\tilde\Phi(q)$ and $\theta(q)$ is real valued function. 
By following the change of the basis,
the connection field $\tilde{\cal A}'_{\mu}(q)$ and the field strength $\tilde{\cal F}'_{\mu\nu}(q)$ are defined in the same way as the above.
Also by the metric compatibility condition for the constant metric, the U(1) gauge boson field ${\cal B}'_{\mu}(q)$ is introduced, which is related to
$\tilde{\cal A}'_{\mu}(q)$ as $\tilde{\cal A}'_{\mu}(q)=ig{\cal B}'_\mu(q)$. The transformation rule of the  
U(1) gauge boson field ${\cal B}_{\mu}(q)$ and the field strength $\tilde{\cal F}_{\mu\nu}(q)$ under 
the (little) gauge transformation is as follows:
\beqa
    {\cal B}'_{\mu}(q)&=&{\cal B}_{\mu}(q)+\frac{1}{g}\del_{\mu}\theta(q),\nonumber\\
     \tilde{\cal F}'_{\mu\nu}(q)&=&\tilde{\cal F}_{\mu\nu}(q)=ig\Big(\del_\mu{\cal B}_{\nu}-\del_{\nu}{\cal B}_{\mu}\Big).
\eeqa

Under the general coordinate transformation $\breve{q}=\breve{q}(q)$, 
the U(1) gauge boson field $\breve{{\cal B}}_{\mu}(\breve{q})$ and the field strength ${\breve{\tilde{\cal F}}_{\mu\nu}}(\breve{q})$ are also defined in the little gauge theory.
It is also shown that they transform as 
\beqa
    &&\breve{{\cal B}}_{\mu}(\breve{q}) ={\Lambda_{\mu}}^{\nu}(q){{\cal B}}_{\nu}({q}),\\
    &&{\breve{\tilde{\cal F}}}_{\mu\nu}(\breve{q}) 
    ={\Lambda_{\mu}}^{\rho}(q){\Lambda_{\nu}}^{\sigma}(q){\tilde{\cal F}}_{\rho\sigma}({q}),
\eeqa 
where ${\Lambda_{\mu}}^{\nu}(q)=\frac{\del q^nu}{\del \breve{q}^\mu}$.
   
By mapping from this little gauge theory to the gauge theory as 
$\be(q)=\varphi(q)\tilde\be(q)$, the connection field ${\cal A}_{\mu}(q)$ in the gauge theory is tamed 
as 
\beq
    {\cal A}_{\mu}(q)=ig{\cal B}_{\mu}(q)+\del_{\mu}\varphi(q)
\eeq 
in terms of Hermite field ${\cal B}_{\mu}(q)$. 
And it turns out that the little gauge theory plays important role to determine
the form of the connection field in the gauge theory.

\section{Dirac fermion field theory}

In this section, the Dirac fermion field theory is considered based on the little gauge theory.
This example is the case of the complex target manifold
equipped with an indefinite inner product. 
 
Supposed that the Dirac fermion field $\vPsi$ on the 4 dimensional space-time as the physical reality.
For a choice of the space-time parametrisation $q^\mu$, $\vPsi$ is assumed as the function of $q^\mu$.
We choose the basis vector $\be_{I}(q)$ of $\vPsi(q)$ so that the metric of the spinor inner product satisfies the condition
\beq
    (\be_{I}(q),\be_{J}(q))=(\gamma_0)_{IJ},\label{constantmetsp}
\eeq
where $\gamma_a (a=0,1,2,3)$ is Dirac gamma matrix, and the small index $a$ and 
the capital index $I$ stand for the Lorezian index and the spinor component index, respectively.  
Note that the metric $\gamma_0$ is the constant metric. 
Then the spinor inner product for the two fermions $\vPsi_1(q)$ and $\vPsi_2(q)$ is given by
\beq
    (\vPsi_1(q),\vPsi_2(q))=(\gamma_0)_{IJ}{(\Psi_1)^\dag}^I(q)\Psi^J_2(q)={\overline{\Psi_1}(q)}\Psi_2(q).
\eeq

The pull back differential of $\be_{I}(q)$ is described by the little connection field
\beq    
    \displaystyle{\partial_\mu}\be_{I}(q)={{{\cal A}_\mu}^J}_{I}(q)\be_{J}(q).
\eeq
The little connection field has the $4\times4\times4=64$ components.
It is decomposed as the following form in terms of the gamma matrix $\gamma_{a}$:
\beqa
 {{{\cal A}_\mu}^J}_{I}(q)={\cal A}_\mu(q) {(1\!\!1)^J}_I+{{\cal E}_\mu}^a(q) {(\gamma_a)^J}_I+{{\cal B}_\mu}^{ab}(q){(\sigma_{ab})^J}_I\nonumber\\
 \hspace{3cm}+{{\cal A}^{(5)}}_\mu(q) {(\gamma_5)^J}_I+{{{\tilde{\cal E}}_\mu}}\ ^a (q){(\gamma_a\gamma_5)^J}_I,
\eeqa
where $1\!\!1$ is the unit matrix, $\displaystyle\sigma_{ab}=\frac{i}{2}[\gamma_a,\gamma_b]$ 
and $\gamma_5=-i\gamma_0\gamma_1\gamma_2\gamma_3$. 
The metric compatibility condition for the constant metric Eq.(\ref{constantmetsp}) leads to the following equation:
\beqa
 &&({\cal A}^\dag_{\mu}(q)+{\cal A}_{\mu}(q))1\!\!1+({{\cal E}^\dag}{{\ }_{\mu}}^a(q)+{{\cal E}_{\mu}}^a(q))\gamma_a
 +({{\cal B}^\dag}{\ }_{\mu}^{ab}(q)+{{\cal B}_{\mu}}^{ab}(q))\sigma_{ab}\nonumber\\
 &&\hspace{3cm}+ (-{{\cal A}^{(5)}}^\dag_{\mu}(q)+{{\cal A}^{(5)}}_{\mu}(q))\gamma_5
                         +({\tilde{\cal E}^\dag}{{\ }_{\mu}}^a(q)+{\tilde{\cal E}_{\mu}}^a(q))
                         \gamma_a\gamma_5=0.
\eeqa
Therefore these connection fields ${\cal A}_{\mu}(q), {{\cal E}_{\mu}}^a(q), 
{{\cal B}_{\mu}}^{ab}(q), {\tilde{\cal E}_{\mu}}^a(q)$ 
are anti-Hermite fields and ${{\cal A}^{(5)}}_{\mu}(q)$ is Hermite field. 
We redefine the little connection fields by introducing the Hermite fields 
$A_\mu(q),\ {E_\mu}^a(q),\ {\omega_\mu}^{ab}(q),\ {\tilde{E}_\mu}{ }^a(q),\ {A^{(5)}}_\mu(q)$ 
and the real coupling constants $g_1,\ g_2,\ g_3,\ g_4,\ g_5$ as the followings:
\beqa
    &&{\cal A}_{\mu}(q)=ig_1A_\mu(q),\ 
    {{\cal E}_{\mu}}^a(q)=ig_2{E_\mu}^a(q),\ 
    {{\cal B}_{\mu}}^{ab}(q)=ig_3{\omega_\mu}^{ab}(q),\nonumber\\  
    &&{{\cal A}^{(5)}}_{\mu}(q)=g_4{A^{(5)}}_\mu(q),\ 
    {\tilde{\cal E}_{\mu}}{ }^a(q)=ig_5{\tilde{E}_\mu}{ }^a(q).\
\eeqa

The pull back differential of the fermion reality $\vPsi(q)$ is obtained as
\beqa
    \del_\mu \vPsi(q)&=&\del_\mu\Big(\Psi^I(q)\be_I(q)\Big)\nonumber\\
    &=&\Big(\del_\mu\Psi^I(q)+{{\cal A_\mu}^I}_J\Psi^J(q)\Big)\be_I(q)\nonumber\\
    &=&\Big(\del_\mu\Psi^I(q)+\Big(ig_1 {A}_\mu(q) {(1\!\!1)^I}_J+ig_2{{E}_\mu}^a(q) {(\gamma_a)^I}_J
    +ig_3{{\omega}_\mu}^{ab}(q){(\sigma_{ab})^I}_J\nonumber\\
    &&\hspace{2cm}+g_4{{A}^{(5)}}_\mu(q) {(\gamma_5)^I}_J
    +ig_5{{{\tilde{E}}_\mu}}\ ^a(q) {(\gamma_a\gamma_5)^I}_J\Big)
\Psi^J(q)\Big)\be_I(q).
\eeqa
We can define the generalised covariant derivative matrix ${\cal D}_{\mu}$ as
\beqa
    \del_\mu \vPsi(q)&=&\Big({{{\cal D}_\mu}^I}_J\Psi^J(q)\Big)\be_I(q),
\eeqa
where
\beq
    {\cal D}_{\mu}=(\del_\mu+ig_1 {A}_\mu(q))1\!\!1
     +ig_2{{E}_\mu}^a(q) {\gamma_a}
    +ig_3{{\omega}_\mu}^{ab}(q){\sigma_{ab}}
    +g_4{{A}^{(5)}}_\mu(q) {\gamma_5}+ig_5{{{\tilde{E}}_\mu}}(q)\ ^a {\gamma_a\gamma_5}.
\eeq
This generalised covariant derivative is the extension of Sogami's one.

In the little gauge theory, the axial gauge boson with the form $igA^{(5)}(q)\gamma_5$ in the covariant derivative is strongly excluded.
If there exists in the covariant derivative, an anomaly of the metric compatibility condition appears.
All crews of the covariant derivative are interpreted as the gauge bosons.
The field ${A}_\mu(q)$ is the U(1) gauge boson and 
${{\omega}_\mu}^{ab}(q)$ is the spin connection gauge field.
New crews ${{E}_\mu}^a(q)$ and ${{{\tilde{E}}_\mu}}(q)$ are named as 
vielbein gauge boson and axial vielbein gauge boson, respectively.
Note that the vielbein gauge boson and the axial vielbein gauge boson are different 
from the external vielbein. Higgs field is involved in the vielbein gauge boson 
and the axial vielbein gauge boson.
Also, the new crew ${{A}^{(5)}}_\mu(q)$ is named as the pseudo-axial gauge boson.

The field strength ${{\cal F}^J}_{I\mu\nu}(q)$ is defined as
\beq
    {{\cal F}^J}_{I\mu\nu}(q)\be_{J}(q)
    =\Big(\del_{\mu}\del_{\nu}-\del_{\nu}\del_{\mu}\Big)\be_{I}(q).
\eeq
The form of the field strength ${{\cal F}^J}_{I\mu\nu}(q)$ is obtained by the direct calculation.
Here the explicit form is beyond the scope of this article. The details will be given 
by the next article\cite{KK:2017}.

In the little gauge theory, the little gauge transformation ${\Omega_{I}}^{J}(q)$ as 
$\be'_I(q)={\Omega_{I}}^{J}(q)\be_{J}(q)$, which preserve the metric (\ref{constantmetsp}), is given 
by
\beq
    {\Omega_{I}}^{J}(q)
    ={\Big(\exp(i\theta_1(q)1\!\!1+ i\theta_{2}^{a}(q)\gamma_{a} 
    +i\theta_{3}^{ab}(q)\sigma_{ab}+\theta_4(q)\gamma_5+i\theta_5^a(q)\gamma_{a}\gamma_5
    )\Big)_{I}}^J.
\eeq
The transformation is not the unitary transformation. Following the transformation,
the new connection field ${{{\cal A}'_{\mu}}^{J}}_{I}(q)$ is defined, and the connection field
can be expanded by  
\beqa
    {{{\cal A}_{\mu}'}^{J}}_{I}(q)={\cal A}'_\mu(q) {(1\!\!1)^J}_I+{{\cal E}'_\mu}^a(q) {(\gamma_a')^J}_I+{{\cal B}'_\mu}^{ab}(q){(\sigma'_{ab})^J}_I\nonumber\\
 \hspace{3cm}+{{\cal A}'^{(5)}}_\mu(q) {(\gamma'_5)^J}_I+{{{\tilde{\cal E}'}_\mu}}\ ^a (q){(\gamma'_a\gamma'_5)^J}_I,
\eeqa
where $\gamma'_a=\Omega(q)\gamma_a\Omega^{-1}(q), \gamma'_5=\Omega(q)\gamma_5\Omega^{-1}(q)$ and $ \sigma'_{ab}=\Omega(q)\sigma_{ab}\Omega^{-1}(q)$.
The metric compatibility condition 
leads to the introduction of gauge boson fields $A'_\mu(q)$,\ ${E'_\mu}^a(q),$
$\ {\omega'_\mu}^{ab}(q)$, \ ${\tilde{E}'_\mu}{ }^a(q)$,\ ${A'^{(5)}}_\mu(q)$.
From the direct calculation, it is shown that the gauge fields are transformed as
\beqa
    &&A'_\mu(q)=A_\mu(q)+\displaystyle\frac{1}{g_1}\del_{\mu} \theta_1(q),\quad
    {E'_\mu}^a(q)={E_\mu}^a(q)+\displaystyle\frac{1}{g_2}\del_{\mu} \theta^a_2(q),\quad\nonumber\\
    &&{\omega'_\mu}^{ab}(q)={\omega_\mu}^{ab}(q)+\displaystyle\frac{1}{g_3}\del_{\mu} \theta^{ab}_3(q),
    \nonumber\\
    &&{\tilde{E}'_\mu}{ }^a(q)={\tilde{E}_\mu}{ }^a(q)+\displaystyle\frac{1}{g_4}\del_{\mu} \theta^a_4(q)
   ,\quad
    {A'^{(5)}}_\mu(q)={A^{(5)}}_\mu(q)+\displaystyle\frac{1}{g_5}\del_{\mu} \theta_5(q).
\eeqa

In this little gauge theory, all gauge bosons were emergent as the shift of the spinor basis.
The general gauge theory is directly obtained by utilizing the degree of freedom of choices for the basis.
The further research of the Dirac fermion field theory by the little gauge theory 
will be continued in details \cite{KK:2017}.

\section{Discussion}

In this article, we introduced the concept of the little gauge theory.
The heart of the little gauge theory is that it does give the restriction to the connection field 
by the metric compatibility condition. As the result, Hermite gauge fields are appeared 
in the connection field and covariant derivative. Also, the gauge theory is reproduced 
as the map from the little gauge theory. 

As the first example of our construction, Cartan geometry was considered in section 4. 
The metric compatibility condition 
and the external vielbein condition were naturally derived. 
It turns out that some known results for Cartan geometry are reproduced. 
Although the multi-index theory was suppressed in order to avoid complexities in this article,
it is possible to construct extended Cartan geometry with multi-index.

In section 5, 
by the well-known example of the charged boson, it is shown that 
the connection field cannot be tamed in the situation only for the general gauge theory.
By the little gauge theory, it was shown that the little connection field is described 
in terms of the U(1) gauge boson. By the map from the little gauge theory to the gauge theory,
the connection field in the gauge theory was explicitly written by the U(1) gauge boson. 
As the results, the usefullness of the little gauge theory was illustrated.

In section 6, the Dirac fermion field theory was presented.
By following that our question is "Who ordered the basis of the spinor index for the fermion?",
its theory was considered in the framework of the little gauge theory.
The gauge boson fields are emerged from the pull back differential of the basis vector. 
In this framework, it is interpreted that Higgs bosons are involved in the vielbein gauge boson 
and the axial vielbein gauge boson. The invariant action of the Dirac fermion field theory
is obtained by using the recipe \cite{Sogami:1995} 
of the Sogami's generalised covariant derivative, the external
vielbein and Lorentzian metric. In addition, by following his recipe, 
the gauge coupling constants are associated with each other.
And, in the invariant action, there does exist the curvature field made by the spin connection field,
which is different from the little connection field of Cartan geometry.
The details will be appeared in \cite{KK:2017}.
Then the external vielbein is related to the dynamics of the
fermion and the gauge bosons through the action. 
The dark matter and dark energy may be explained in this direction. 

However, our little gauge theory of the Dirac fermion field theory is 
not achieved at the phenomenological level. 
A theory involving all elementary particles should be presented.
In the direction, SO(10) grand unified theory and so on are considered.
As one of other directions, the little gauge theory can be applied to 
the unified description of quarks and leptons in a multi-spinor field formalism 
\cite{Sogami:2015}. New investigations based on the little gauge theory will be opened 
towards to new physics beyond the standard model and the quantum theory of gravity. 

In this article, the connection fields and the gauge fields are introduced,
as our recognition of the space-time is four dimensional manifold.
By the definition, the connection fields are propagated only in the four dimension.
The reason why our recognition of the space-time is four dimensional manifold 
is veiled in this line. It will be clarified by the string theory or the super string theory.

\section*{Acknowledgements}
I would like to express the deepest appreciation to Dr. I. S. Sogami 
for inspiring me to renewed this research and 
giving me constructive comments and warm encouragement.
I would like to thank Dr. N. Ikeda and Dr. K. Hamachi 
for informative and valuable conversations for a lot of my questions.
I also thank my lovely family for a lot of encouragements and supports.  

\section{Appendix}
\subsection{Proof of the first and second Bianchi identities}

In this appendix, the first and second Bianchi identities are proved for the curvature on Cartan geometry.

First, we derive the first Bianchi identity in our formalism.
Starting from the definition of the torsion (\ref{deftorsion}), 
the pull back differential leads to
\beq
 \del_{\alpha}\Big({T_{\beta\gamma}}^\lambda(q)\hbe_{\lambda}(q)\Big)
 =\del_{\alpha}\Big(\del_{\beta}\hbe_{\gamma}(q)-\del_{\gamma}\hbe_{\beta}(q)\Big).\label{Bianki1-1}
\eeq
By using the covariant derivative, the l.h.s. in the above equation (\ref{Bianki1-1}) is written as
\beqa
    \del_\alpha\Big({T_{\beta\gamma}}^\lambda(q)\hbe_{\lambda}(q)\Big)
    &=&\Big(\del_{\alpha}{T_{\beta\gamma}}^\lambda(q)
            +{\Gamma^\lambda}_{\rho\alpha}(q){T_{\beta\gamma}}^\rho(q)\Big)\hbe_{\lambda}(q)\nonumber\\
     &=&\Big(\nabla_{\alpha}{T_{\beta\gamma}}^\lambda(q)
            +{\Gamma^\rho}_{\beta\alpha}(q){T_{\rho\gamma}}^\lambda(q)
            +{\Gamma^\rho}_{\gamma\alpha}(q){T_{\beta\rho}}^\lambda(q)\Big)\hbe_{\lambda}(q) 
\eeqa
Taking the circular summation of Eq.(\ref{Bianki1-1}) over the three indices $\alpha,\beta,\gamma$, the l.h.s. is
\beqa
  \displaystyle\sum_{(\alpha\beta\gamma)}\del_\alpha\Big({T_{\beta\gamma}}^\lambda(q)\hbe_{\lambda}(q)\Big)
  &=& \displaystyle\sum_{(\alpha\beta\gamma)}\Big(\nabla_{\alpha}{T_{\beta\gamma}}^\lambda(q)
            +{\Gamma^\rho}_{\beta\alpha}(q){T_{\rho\gamma}}^\lambda(q)
            +{\Gamma^\rho}_{\gamma\alpha}(q){T_{\beta\rho}}^\lambda(q)\Big)\hbe_{\lambda}(q) \nonumber\\
  &=& \displaystyle\sum_{(\alpha\beta\gamma)}\Big(\nabla_{\alpha}{T_{\beta\gamma}}^\lambda(q)
  +{T_{\alpha\beta}}^\rho(q){T_{\rho\gamma}}^\lambda(q)\Big)\hbe_{\lambda}(q),
\eeqa
where $\displaystyle\sum_{(\alpha\beta\gamma)}$ stands for the circular summation,
and the r.h.s is  
\beqa
    \displaystyle\sum_{(\alpha\beta\gamma)}\del_{\alpha}\Big(\del_{\beta}\hbe_{\gamma}(q)
    -\del_{\gamma}\hbe_{\beta}(q)\Big)
    &=&\del_{\alpha}\del_{\beta}\hbe_{\gamma}(q)
    -\del_{\alpha}\del_{\gamma}\hbe_{\beta}(q)\nonumber\\
    &&\hspace{0.5cm}+\del_{\beta}\del_{\gamma}\hbe_{\alpha}(q)
    -\del_{\beta}\del_{\alpha}\hbe_{\gamma}(q)
    +\del_{\gamma}\del_{\alpha}\hbe_{\beta}(q)
    -\del_{\gamma}\del_{\beta}\hbe_{\alpha}(q)\nonumber\\
    &=&\displaystyle\sum_{(\alpha\beta\gamma)}{R^\lambda}_{\alpha\beta\gamma}(q)\hbe_{\lambda}(q)
\eeqa
Thus the first Bianchi identity is obtained as
\beq
   \displaystyle\sum_{(\alpha\beta\gamma)}\Big(\nabla_{\alpha}{T_{\beta\gamma}}^\lambda(q)
  +{T_{\alpha\beta}}^\rho(q){T_{\rho\gamma}}^\lambda\Big)
  =\sum_{(\alpha\beta\gamma)}{R^\lambda}_{\alpha\beta\gamma}(q).
\eeq

In the next, we derive the second Bianchi identity by the similar calculation to the first Bianchi identity. By taking the pull back differential for the definition of the curvature (\ref{defcur}) and 
the circular summation over the three indices $\alpha,\beta,\gamma$, we obtain the equation
\beq
    \displaystyle\sum_{(\alpha\beta\gamma)}\del_{\alpha}\Big({R^\rho}_{\sigma\beta\gamma}(q)\hbe_{\rho}(q)\Big)=\sum_{(\alpha\beta\gamma)}\del_{\alpha}\Big(\del_\beta\del_\gamma-\del_\gamma\del_\beta\Big)\hbe_{\sigma}(q) \label{B2-1}.
\eeq
By using the covariant derivative, the l.h.s of the above Eq. (\ref{B2-1}) is
\beqa
 \hspace{-1.1cm}\displaystyle\sum_{(\alpha\beta\gamma)}\del_{\alpha}\Big({R^\rho}_{\sigma\beta\gamma}(q)\hbe_{\rho}(q)\Big)
    &\!\!\!=\!\!\!&\sum_{(\alpha\beta\gamma)}\Big(\del_{\alpha}{R^\rho}_{\sigma\beta\gamma}(q)\Big)\hbe_{\rho}(q)
    +{R^\rho}_{\sigma\beta\gamma}(q)\Big(\del_{\alpha}\hbe_{\rho}(q)\Big)\nonumber\\
    &=&\displaystyle\sum_{(\alpha\beta\gamma)}
    \Big(\nabla_{\alpha}{R^{\rho}}_{\sigma\beta\gamma}(q)
    +{\Gamma^\mu}_{\sigma\alpha}(q){R^{\rho}}_{\mu\beta\gamma}(q)\nonumber\\
    &&\hspace{0.8cm}+{\Gamma^\mu}_{\beta\alpha}(q){R^{\rho}}_{\sigma\mu\gamma}(q)
    +{\Gamma^\mu}_{\gamma\alpha}(q){R^{\rho}}_{\sigma\beta\mu}(q)\Big)\hbe_{\rho}(q)\nonumber\\
    &&\hspace*{-1.8cm}=\displaystyle\sum_{(\alpha\beta\gamma)}
    \Big(\nabla_{\alpha}{R^{\rho}}_{\sigma\beta\gamma}(q)
    -{T_{\beta\gamma}}^\mu(q){R^{\rho}}_{\sigma\alpha\mu}(q)
    +{\Gamma^{\mu}}_{\sigma\alpha}(q){R^{\rho}}_{\mu\beta\gamma}(q)\Big)\hbe_{\rho}(q).
\eeqa
On the other hand,  the r.h.s of Eq. (\ref{B2-1}) is
\beqa
&&\displaystyle\sum_{(\alpha\beta\gamma)}\del_{\alpha}\Big(\del_\beta\del_\gamma-\del_\gamma\del_\beta\Big)\hbe_{\sigma}(q)\nonumber\\
&&=\Big(\del_{\alpha}\del_{\beta}\del_{\gamma}-\del_{\alpha}\del_{\gamma}\del_{\beta}
+\del_{\beta}\del_{\gamma}\del_{\alpha}-\del_{\beta}\del_{\alpha}\del_{\gamma}
+\del_{\gamma}\del_{\alpha}\del_{\beta}-\del_{\gamma}\del_{\beta}\del_{\alpha}
\Big)\hbe_{\sigma}(q)\nonumber\\
&&=\sum_{(\alpha\beta\gamma)}\Big(\del_\beta\del_\gamma-\del_\gamma\del_\beta\Big)\del_\alpha\hbe_{\sigma}(q)\nonumber\\
&&=\sum_{(\alpha\beta\gamma)}\Big(\del_\beta\del_\gamma-\del_\gamma\del_\beta\Big){\Gamma^{\rho}}_{\sigma\alpha}(q)\hbe_{\sigma}(q)\nonumber\\
&&=\sum_{(\alpha\beta\gamma)}{\Gamma^{\rho}}_{\sigma\alpha}(q)\Big(\del_\beta\del_\gamma-\del_\gamma\del_\beta\Big)\hbe_{\sigma}(q)\nonumber\\
&&=\sum_{(\alpha\beta\gamma)}{\Gamma^{\rho}}_{\sigma\alpha}(q){R^{\rho}}_{\mu\beta\gamma}(q)\hbe_{\sigma}(q),
\eeqa
where $\Big(\del_\beta\del_\gamma-\del_\gamma\del_\beta\Big){\Gamma^{\rho}}_{\sigma\alpha}(q)\hbe_{\sigma}(q)={\Gamma^{\rho}}_{\sigma\alpha}(q)\Big(\del_\beta\del_\gamma-\del_\gamma\del_\beta\Big)\hbe_{\sigma}(q)$
was used. As these results, we obtain the second Bianchi identity
\beq
    \sum_{(\alpha\beta\gamma)}
    \Big(\nabla_{\alpha}{R^{\rho}}_{\sigma\beta\gamma}(q)
    -{T_{\beta\gamma}}^\mu(q){R^{\rho}}_{\sigma\alpha\mu}(q)\Big)=0.
\eeq


\end{document}